# Lithium Diffusion in Graphitic Carbon

Kristin Persson,[1,2,*,+] Vijay A. Sethuraman,[1,3,*] Laurence J. Hardwick,[1,4,*] Yoyo Hinuma,[2,5] Ying Shirley Meng,[2,5] Anton van der Ven,[6] Venkat Srinivasan,[1] Robert Kostecki,[1] Gerbrand Ceder[2]

[1]*Lawrence Berkeley National Laboratory, 1 Cyclotron Rd., Berkeley CA 94720, USA*
[2]*Massachusetts Institute of Technology, 77 Mass Ave., Cambridge MA 02139, USA*
[3]*Brown University, 182 Hope Street, Providence RI 02906, USA*
[4]*University of St Andrews, North Haugh, St Andrews, Fife KY16 9ST, Scotland, UK*
[5]*University of California San Diego, Atkinson Hall 2703 La Jolla, CA 92093 USA.*
[6]*University of Michigan, 2300 Hayward St., Ann Arbor, MI 48109, USA*

[*] - contributed equally to this work

[+]**Corresponding Author:** Kristin Persson, e-mail: kapersson@lbl.gov

Graphitic carbon is currently considered the state-of-the-art material for the negative electrode in lithium-ion cells, mainly due to its high reversibility and low operating potential. However, carbon anodes exhibit mediocre charge/discharge rate performance, which contributes to severe transport-induced surface-structural damage upon prolonged cycling, and limits the lifetime of the cell. Lithium bulk diffusion in graphitic carbon is not yet completely understood, partly due to the complexity of measuring bulk transport properties in finite-sized, non-isotropic particles. To solve this problem for graphite, we use the Devanathan-Stachurski electrochemical methodology combined with *ab-initio* computations to deconvolute, and quantify the mechanism of lithium-ion diffusion in highly oriented pyrolytic graphite (HOPG). The results reveal inherent high lithium-ion diffusivity in the direction parallel to the graphene plane (*ca.* $10^{-7}$ - $10^{-6}$ cm$^2$ s$^{-1}$), as compared to sluggish lithium-ion transport along grain boundaries (*ca.* $10^{-11}$ cm$^2$ s$^{-1}$), indicating the possibility of rational design of carbonaceous materials and composite electrodes with very high rate capability.

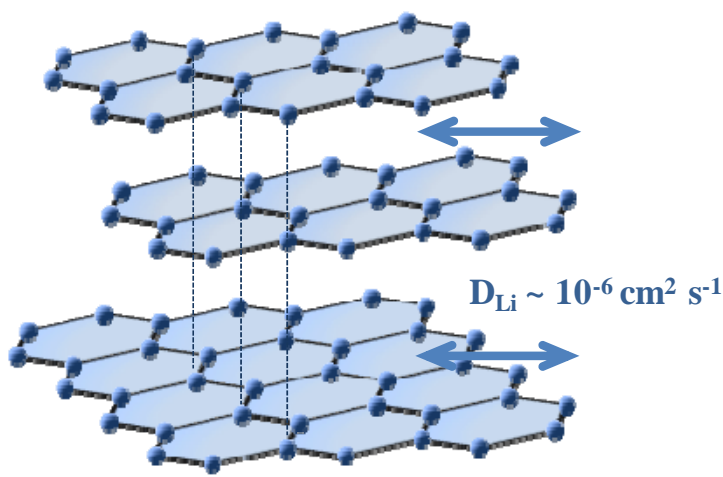

$D_{Li}$ ~ $10^{-6}$ cm$^2$ s$^{-1}$

While commercial lithium-ion batteries can consist of multiple cathode chemistries, a vast majority of them use graphitic carbon as the negative-electrode material because of its low cost, low operating potential, high capacity, high reversibility, and remarkable structural and interfacial stability.[1] The use of graphitic electrodes as ion-intercalation negative-electrode hosts for rechargeable electrochemical power sources was suggested first by Rudorff and Hofmann[2] in 1938, and many scientists have subsequently investigated them.[3] Lithium diffusion in graphitic carbon is not yet completely understood due to a lack of reliable theoretical and experimental methods. Impedance spectroscopy,[5-8] potentiostatic intermittent titration technique (PITT),[9] and standard electrochemical methods[9,10] have been used to gauge the diffusion coefficient in different types of graphitized carbons in composite electrodes, and to determine their overall rate performance in lithium-ion systems. However, the electrochemical response of a composite electrode consists of multiple components, and in principle fails to resolve the highly inherent anisotropic nature of lithium diffusion in graphitic carbon. Thus, the basic electrochemical properties of graphite becomes convoluted with the parameters of mass, and charge transfer involving particle contact resistances, surface films, and side reactions. Moreover, the analysis of the experimental data is extremely complicated, and consequently, the lithium-ion transport rates reported for various types of composite-graphite electrode architectures vary in the literature from $10^{-6}$ to $10^{-16}$ cm$^2$ s$^{-1}$.[5-12]

In this letter, we present a combination of electrochemical measurements of lithium-ion permeation, and first-principles calculations to clarify and quantify lithium-ion diffusion in HOPG. The objective of this work is to (i) determine the diffusion paths and lithium-ion transport parameters in graphite, and (ii) provide rational guidelines for design and synthesis of high-rate graphitic materials.

In order to directly measure Li diffusion in graphite, an HOPG foil was used as a membrane in a Devanathan-Stachurski type two-compartment cell.[13] The HOPG used is made of single-crystal graphitic cuboids (*i.e.*, graphene domains) with an angular spread of the c-axes of the crystallites less than 1 degree. The graphene basal-planes are exposed at the surface of the HOPG foil whereas the plane edges are at the foil perimeters. The HOPG membrane served as a common working electrode for both compartments "A" and "B" (see Figures 1 and 2), which were filled with 1.2 M LiPF$_6$ in EC:EMC (1:1) electrolyte and equipped with two sets of metallic lithium reference and counter electrodes. Further details of the materials and cell setup are described in the supporting material. Li permeation measurements were carried out with two types of HOPG membranes. In the first experiment, a thin and flat 20 μm thick HOPG membrane was used (see inset, Figure 1) and in the second case, a 3 mm thick HOPG membrane with two partial and overlapping holes (diameter = 1 mm, 0.5 mm apart) carefully drilled from the opposing sides of the membrane was used (see inset, Figure 2). The objectives of these experiments were to measure, respectively, the lithium-ion diffusion perpendicular to the graphene layers (*i.e.*, diffusion across the basal plane, see Figure 1) and parallel to them (*i.e.*, diffusion along the basal plane, see Figure 2).

The surface of the HOPG membrane in compartment A was polarized galvanostatically with an applied current of 25 μA cm$^{-2}$, which enabled constant rate lithium insertion into the graphite, while the other side of the membrane in compartment B was maintained at sufficiently high anodic potential (3 V *vs.* Li/Li$^+$) to immediately oxidize all Li that moved across and appeared at the B-side of the HOPG membrane. The time delay between the lithium insertion at the A-side and the anodic current response at the B-side of the membrane constitutes a direct measurement of the transport rate of lithium through HOPG. While the electrochemical processes at the surface of the HOPG membrane in compartment A also involve formation of the solid-electrolyte interphase (SEI), its effect on the current response in compartment B is negligible.

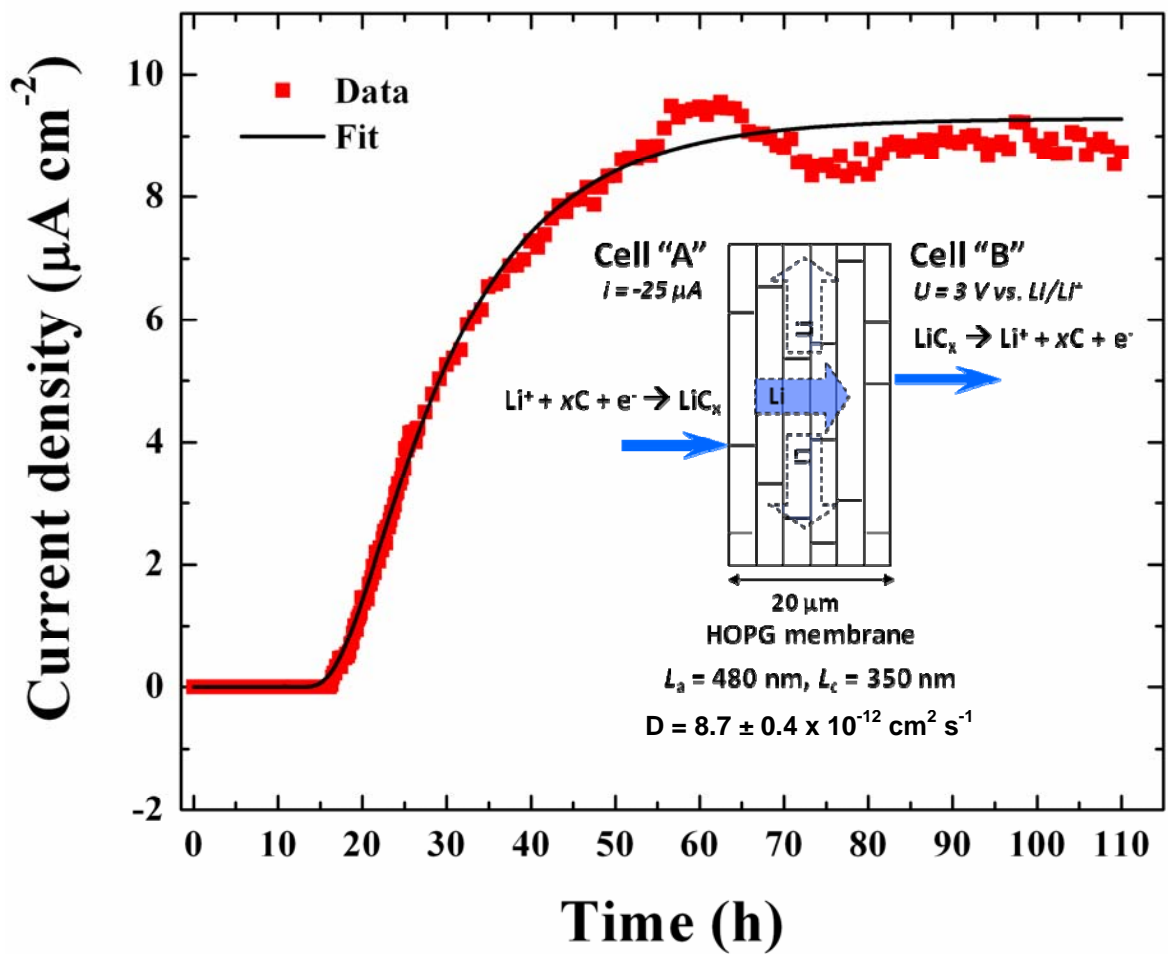

*Figure 1. The chronoamperometric response obtained on the HOPG membrane on Cell B in the basal-plane configuration. The current corresponds to the de-intercalative flux of lithium ions from the HOPG membrane into the electrolyte. Inset: Schematic showing the diffusion pathway of lithium ions in the basal-plane configuration. A constant intercalative flux of lithium ions was imposed on the electrolyte/HOPG interface in compartment A while a constant potential of 3 V vs. Li/Li$^+$ was maintained on the HOPG/electrolyte interface in compartment B.*

The anodic current response for the thin HOPG membrane is shown in Figure 1. Because only basal planes were exposed to the electrolyte, lithium ions could diffuse through HOPG solely between the graphene crystallites and then toward the B-side along the grain boundaries. A small amount of Li$^+$ could also penetrate into HOPG structure via point defects and step edges at the surface of the graphene planes but their contribution to the transport mechanism in crystalline graphite is insignificant. The current threshold in compartment B is observed after *ca.* 17 hours, followed by a steep rise to reach a plateau after *ca.* 80 hours. Interestingly, only a small fraction of the Li$^+$ inserted in the HOPG membrane was detected on the B side. The missing lithium most likely diffused into the part of the membrane that was not exposed to the electrolyte. The length and breadth of the HOPG membrane are much larger (few centimeters) than its thickness (~60 microns). Because of this, it would take a very long time for all the lithium (that was intercalated into the HOPG) to de-intercalate on side B. Since the rise-time (of the chorono-ameperometric response on side B), and the steady-value of the current were of interest, the experiment was not performed for such large timescales.

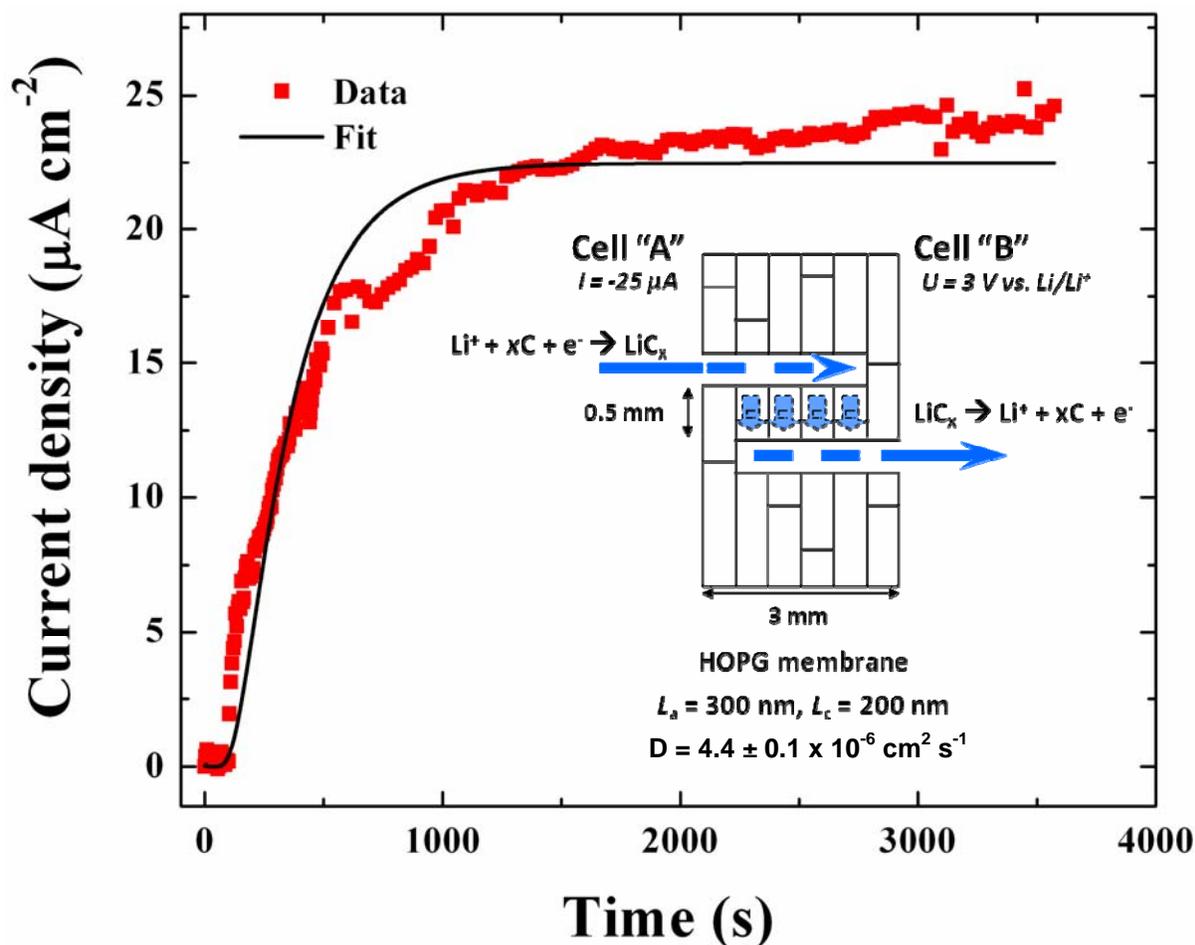

*Figure 2. The chronoamperometric response obtained on the HOPG membrane in compartment B in the edge-plane configuration. The current corresponds to the de-intercalative flux of lithium-ions from the HOPG membrane into the electrolyte. Inset: Schematic showing the diffusion pathway of lithium ions in the edge plane configuration. A constant intercalative-flux of lithium ions was imposed on the electrolyte/HOPG interface in compartment A while a constant potential of 3 V vs. Li/Li$^+$ was maintained on the HOPG/electrolyte interface in compartment B.*

The thick HOPG membrane with two partial and overlapping holes offered two types of entry sites for lithium *i.e.*, interplanar, and graphene domain boundaries (Figure 2, inset). In this case, the current response in compartment B was observed after only 170 seconds to reach a plateau after *ca.* 20 minutes (Figure 2). The observed nearly 100 % coulombic efficiency of redox processes in A and B compartments indicates that almost all Li$^+$ inserted in this HOPG membrane reached the other side of the membrane.

Using the chrono-amperometric response obtained in compartment B in both experiments together with Fick's second law for the diffusion equation, along with the appropriate initial and boundary conditions for the experimental setup (see supporting material), the lithium-ion diffusion coefficients in the direction perpendicular and parallel to the graphene planes were estimated. Thus, the average diffusion coefficient of lithium-ions in graphite was determined as 8.7 ($\pm$ 0.4) x 10$^{-12}$ cm$^2$ s$^{-1}$ in the direction perpendicular to graphene planes in HOPG, and 4.4 ($\pm$ 0.1) x10$^{-6}$ cm$^2$ s$^{-1}$ in the direction parallel to graphene planes. The respective error-margins were obtained by setting the 95% confidence on the estimated parameters during the parameter estimation routine.

There have been several *ab initio* studies on the lithium-graphite system[14-17], but the chemical diffusion coefficient of lithium as a function of concentration has not been calculated from first principles. Verbrugge *et al.*[10] treated the Li diffusivity in graphite within the 1-D continuum transport framework and Toyoura *et al.*[16,17] recently calculated the Li diffusivity in ordered $LiC_6$ assuming a single vacancy or interstitial diffusion mechanism.

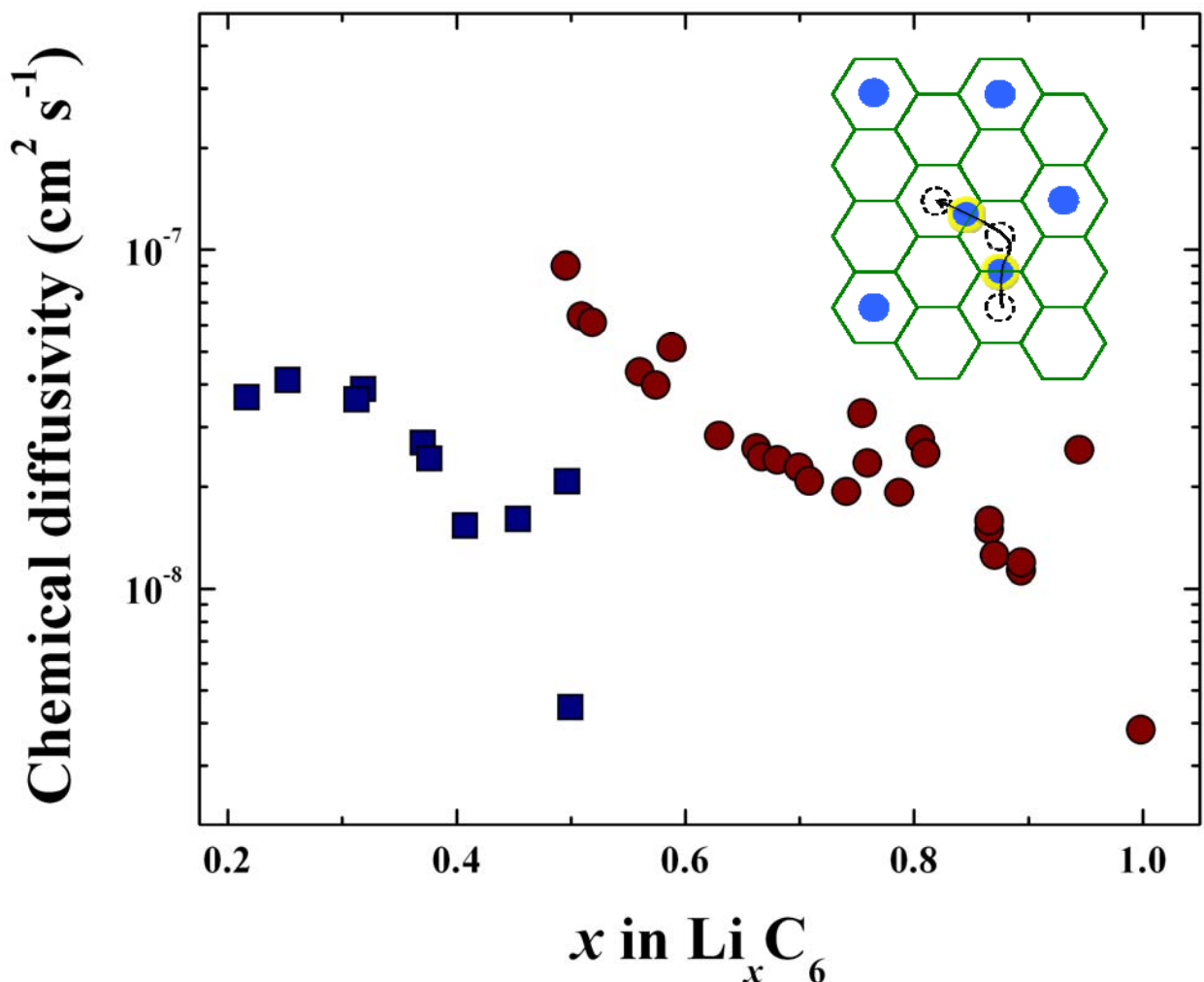

*Figure 3: The chemical diffusivity for the stage II (blue squares) and stage I (red circles) phases in the Li-graphite system, obtained from kinetic Monte Carlo simulations based on first-principles calculations.[28] The inset shows the Li (blue circles) path on the honeycomb graphite lattice which was used to calculate the Li migration barriers.*

In order to calculate a general concentration-dependent lithium diffusivity in graphite from first principles, without making any assumptions regarding the diffusion mechanism, we use the cluster expansion method[18] to model partially disordered states at finite temperatures. Such an approach has, for example, been used successfully to elucidate the lithium-ion diffusion in $Li_xCoO_2$ and $Li_xTiS_2$.[19,20] It has been shown that the lithium-ion intercalation in graphite occurs in stages, where stage *n* contains (*n* - 1) empty layers between each lithium-filled layer.[21,22] We chose to focus our investigation on stage I and stage II compounds, as these phases dominate the lithium-graphite phase diagram[23] (as well as the concentrations that resulted inside the HOPG membrane in both experiments) and will therefore provide a good representation of lithium motion in graphite as a function of concentration. Thus, the energies of sixty-three stage I and stage II structures of different lithium concentrations and Li arrangements were calculated through the generalized gradient approximation [24] to density functional theory as implemented in the Vienna Ab Initio Simulation Package (VASP).[25] Projected augmented wave pseudopotentials[26, 27] were used, with an energy cutoff of 400 eV. Both internal coordinates and unit cell lattice parameters were fully relaxed and the Brillouin zones were sampled with a gamma-centered mesh so that the energy convergence with respect to the $k$-point sampling was better than 5 meV/6C. Further details and motivations regarding the calculations can be found in the Supporting material. The resulting energies were used to fit a cluster expansion which was employed to calculate the lithium-graphite phase diagram through Monte Carlo simulations and benchmarked against experiments.[23] The full details of these calculations and extended analyses of the lithium-graphite thermodynamics will be published elsewhere.[28]

Kinetic Monte Carlo simulations were employed to calculate lithium diffusion coefficients as a function of lithium concentration in stage I and stage II compounds. In defect-free graphite, lithium motion is restricted to two-dimensional diffusion because lithium hopping between layers through a carbon honeycomb is energetically extremely unfavorable.[29] According to transition state theory, the frequency with which lithium ions move to vacant neighboring sites is expressed as:

$$\Gamma = \nu^* \exp(-\Delta E_h / k_B T) \qquad [1]$$

where $\Delta E_h$ is the difference between the energy at an activated state and the initial equilibrium state, and $\nu^*$ is an effective vibrational frequency, here taken as $1 \times 10^{13}$ s$^{-1}$.[17] The location and energy of the activated states, calculated separately for stage I and II, were determined by the nudged-elastic-band method.[30] The migration barriers were combined with the cluster expansion for the lithium–vacancy configuration energy[31] in graphite and used to construct activation barriers in kinetic Monte Carlo simulations to calculate the lithium diffusion coefficients. We have chosen to show the diffusivity results for stage I (II) for the concentration regions where experiments[5] indicate average intra-layer spacings characteristic of stage I (II). As can be seen in Figure 3, the chemical diffusivity in stage I (II) decreases as a function of increasing in-plane Li concentration. This is a direct result of repulsive lithium–lithium interactions, which inhibit lithium mobility at higher lithium content. Apart from the ordering effects, i.e., some fluctuations and a sharp decrease in diffusivity at $x = 0.5$ and $x = 1.0$, there are no other significant features in the diffusivity trend with concentration. Most importantly, we find overall very fast intra-layer lithium diffusion in bulk graphite, ca. $10^{-7}$ cm$^2$ s$^{-1}$ at room temperature. This is in very good agreement with the experimental findings for diffusion parallel to graphene planes in HOPG, at $4.4 (\pm 0.1) \times 10^{-6}$ cm$^2$ s$^{-1}$ and gives further evidence for the extremely high mobility of Li between the layers of graphite.

To conclude, we have studied, by highly controlled experiments combined with first principles calculations, lithium diffusivity in HOPG as a function of transport direction. The diffusivities obtained from these efforts show remarkable agreement between experiments and calculations, and clearly indicate that the lithium diffusion in graphite is several orders of magnitude faster between graphene planes than along the grain boundaries or in the direction perpendicular to the graphene sheets. While this is perhaps intuitively not surprising, it has several important implications. Firstly, it provides a physical explanation for the very wide range of lithium diffusivity data that is reported in the literature for different degrees of graphitized carbons. Traditionally, this inconsistency has been attributed to the planar surface models, used in analysis of diffusivity experiments, which can differ significantly with the actual electrochemical interface area.[12] While this uncertainty about the active area undoubtedly contributes to the variation in the literature, the results presented in this paper also indicate that the lithium diffusivity in any graphitic carbon will depend critically on the size of graphitized domains as well as its orientation relevant to the intercalative/de-intercalative flux. Secondly, the findings have immediate implications for potential rational design of carbonaceous materials for high-rate anodes in lithium-ion batteries. It is well known that graphite anodes suffer severe transport-induced surface-structural damage upon prolonged cycling (especially at high rates and at elevated temperatures) in rechargeable lithium-ion batteries.[32-35] While well-controlled structure and performance-oriented design of cathode materials have recently been explored,[36] there are currently no guidelines for designing carbon-based electrode architectures for lithium ion batteries, especially for high-power applications.

Assuming a design which efficiently utilizes the fast in-plane lithium diffusivity of $10^{-7}$ $cm^2$ $s^{-1}$, graphitized natural graphite (MCMB) with typical crystalline domain sizes around 45 nm could be intercalated/deintercalated in less than 0.2 milli-seconds. Such rate would compete with the fastest-rate cathode materials seen to date.[37] For example, a rate-promoting design could potentially be achieved by creating graphite particles with radially-aligned crystallites where the graphene planes are parallel to the each other in radial direction. An anode material with this kind of structural alignment should, by construction, exhibit very little disordering which minimizes the irreversible capacity loss of lithium to solid electrolyte interphase (SEI) formation[38] and reduce Li plating which is the main degradation mechanism for Li-ion batteries operating at low temperature. While our work shows that graphitic carbon can in principle be a very high rate anode, and hence enable fast charging batteries, it is important to understand that lithium diffusion in a composite electrode is only as fast as the weakest link in the chain. Hence, electronic and/or ionic transport through the SEI layer, or Li-ion transport through the electrode porosity may also have to be optimized in order to enable high rate anodes.

**Acknowledgements.** Work at the Lawrence Berkeley National Laboratory was supported by the Assistant Secretary for Energy Efficiency and Renewable Energy, Office of Vehicle Technologies of the U.S. Department of Energy, under contract no. DE-AC02-05CH11231. Work at the Massachusetts Institute of Technology was supported by Ford Motor Company under Grant Number 014502-010. AVDV acknowledges support from NSF under grant number DMR 0748516.

**Supporting Information Available.** Experimental setup, Diffusion equations and the estimation of transport parameters, First-principles calculations. This information is available free of charge via the internet at http://pubs.acs.org.